\documentstyle[aps,tighten,floats,epsf]{revtex}
\begin{document}
\renewcommand
\baselinestretch{2}
\large
\title{\huge Stochastic Resonance in Washboard Potentials}
\author{Debasis Dan$^{1}$\cite{dannny}, Mangal. C. Mahato$^{2}$, A. M. Jayannavar$^{1}$
\cite{jayan}}
\address{1. Institute of Physics, Sachivalaya Marg, Bhubaneswar 751005, India.\\
   2.  Department of Physics, Guru Ghasidas University, Bilaspur 495009, India} 
\maketitle
\vskip 0.3in
\begin{abstract}
\begin{flushleft}
{\bf ABSTRACT}
\vskip 0.2in

\hspace{0.6in}	We study the mobility of an overdamped particle in a periodic potential tilted by a
constant force. The mobility exhibits a stochastic resonance in 
inhomogeneous systems with space dependent friction coefficient. The result
indicates that the presence of oscillating external field is not essential
for the observability of stochastic resonance, at least in the inhomogenous
medium.
\end{flushleft}
\end{abstract}
\section{Introduction }
	Stochastic Resonance( S.R) is an important phenomena with considerable 
implications in all branches of science \cite{Gam}. The enhanced response of a nonlinear system to input signals at the expense of and as a function of noise is termed as
Stochastic Resonance. It is generally accepted that SR can be observed provided certain essential conditions are fulfilled.
 Attempts are being continually made to reduce the number and stringency
 of these constraints for the realization of the phenomenon.
 The simplest and the minimal ( currently accepted) conditions under which conventional SR can
 be observed are, the presence of a) a bistable system,  b) a tunable Gaussian
 white noise and c) a time varying periodic signal (force).
 Recently, Hu \cite{Hu} suggested that the last ingredient may hopefully be replaced
 by a constant force and, by implication, SR may be observed, e.g, in the drift
 velocity ( mobility) of an overdamped Brownian particle in a tilted periodic potential as a function of noise strength.
 Unfortunately the suggestion was proved to be incorrect \cite{Git,Cas,Mar}.
 However, Marchesoni, \cite{Mar} by analyzing the work of Risken \cite{Ris}, observed that
 SR can be observed in the drift velocity of Brownian particles in a tilted
 periodic (washboard) potential only in the underdamped situation where
 friction acts as surrogate to the external periodic drive.
 In the present work we show that SR can be observed in the mobility
 of even an overdamped Brownian particle in a tilted periodic potential (without
 the presence of oscillating field) but in the presence of a space-dependent
 periodic friction coefficient( i.e., in an inhomogeneous medium).

 The space dependence of friction coefficient does not affect
 the equilibrium properties such as the equilibrium probability distribution.
 However, it does affect the dynamical (nonequilibrium) properties of the
 system( such as the relaxation rates). The space dependence of friction $\eta(q)$  can be microscopically
modeled through the nonlinear( space-dependent)
 coupling between the particle degrees of freedom and the thermal bath(
 characterized by its equilibrium temperature) \cite{Jay,Mah,Bao}.
 In this work we find that the mobility of overdamped particles in a sinusoidal
 potential subject to a sinusoidal friction coefficient 
 but with a phase difference $\phi$ 
 shows a peak as a function of noise strength( temperature of the bath) 
 in the presence of only a constant force field $F$.
  By properly choosing $\phi$  we obtain this SR behaviour in the
mobility( defined as drift velocity
 divided by $F$) even for a small constant external force field ( when the
 barrier for the particle motion in the tilted periodic potential remains finite).

The motion of an overdamped particle, in a periodic potential $V(q)$ and subjected
 to a space-dependent friction coefficient $\eta(q)$  and an additional
constant force $F$, at temperature $T$, is described by the
 Fokker-Planck equation \cite{Jay,Mah,Kam,Kam2}.
\begin{equation}
 \frac{\partial P}{\partial t}  =  \frac{ \partial}{\partial q}  \frac{1}{\eta (q)}[k_{B}T \frac {\partial P}{\partial q}  +  (V^{\prime}(q)-F)P]
\end{equation}
One can calculate \cite{Ris,Mah,But} the probability current $j$, for the potential function
$V(q)$ with $V(q+2\pi)=V(q)$, as
\begin{equation}
 j  =  \frac{k_{B}T(1-exp(\delta))}{\int_{0}^{2\pi}exp(-\psi (y))dy \int_ {y}^{y+2\pi}\eta (x)exp(\psi (x))dx} ,
\end{equation}
where
\begin{eqnarray*}
\psi (q)=\int ^{q}\frac{V^{\prime }(x)-F}{k_{B}T}dx \\
=\frac{V(q)-Fq}{k_{B}T}
\end{eqnarray*}
and $\delta =\psi (q)-\psi (q+2\pi )=\frac{2\pi F}{k_{B}T}$. The mobility( defined \cite{Ris} as the ratio of the drift velocity,
$\left\langle\frac{dq}{dt}\right\rangle = 2\pi j$ divided by the applied force $F$),
$ \mu =\frac{\left\langle \frac{dq}{dt}\right\rangle }{F}=\frac{2\pi j}{F}$. We
take $V(q)=-\sin(q)$ and $\eta(q)=\eta _{0}(1-\lambda \sin (q+\phi ))$, with
$0\leq \lambda < 1$. Clearly, $j\rightarrow 0$ as $F\rightarrow 0$, but the
mobility $\mu$ remains finite as $F \rightarrow 0$, for nonzero temperature $T$.
However, for $F\leq 1$ as $T\rightarrow 0$, $\mu \rightarrow 0$. Also, as
$F, T\rightarrow  \infty, \mu \rightarrow  \frac{1}{\eta _{0}}$. However, for 
given $\lambda$ and $\phi$, in order to find the variation of $\mu$ at intermediate
$T$ and $F$, one needs to evaluate the double integral in the denominator of eqn.
(2) numerically \cite{Alan}.

\section{The Results }

The variation of mobility $\mu$ in the parameter space of $( T, F, \lambda, \phi)$
provides a very rich structure where the phase lag $\phi$ plays an important role.
However, in this paper we report only the variation of $\mu$  with temperature
( noise strength) $T$ at a few carefully selected values of $(F, \lambda, \phi)$
in order to highlight the observability of S.R.

From eqn. (2) one can find that, even though $j(F) \neq -j(-F)$, except for phase lag
$\phi = 0$ or $\pi$, but yet $\mu (F,\phi )=\mu (-F,-\phi )$. Therefore, we 
 need to explore only $F > 0$.  Also in order to observe SR  the choice of larger values of 
$\lambda$ is found to be more appropriate. So we take $\lambda =0.9$ 
and explore the full range of  $\phi [0,2\pi ]$. We find that there is a range of values of
$\phi$ within which SR is observed. For example, for $\phi=0.8\pi, \lambda=0.9$,
SR is observed but the peaks are very broad. And for this $\phi$, SR is more
prominent for $F < 1$ than $F > 1$. For smaller $\phi$ it is harder to observe  the peaks as they become still broader.
However, as $\phi$ is increased the peaks become sharper. Fig.1 
shows the mobility $\mu$( in dimensionless units $\eta_{0}\mu$) as a function of $T$( in dimensionless units) for $\phi=0.9\pi$ and $\lambda=0.9$. Here 
the peaks are larger for $F < 1$ than for $F > 1$. Fig. 2, shows the mobility
$\mu$ as a function of $T$ for $\phi = 1.44\pi$, $\lambda = 0.9$. It can be
seen that the peaks are almost flat for $0 < F <1$, but are prominent for the intermediate
temperature range. From Fig. 3, we do not observe S.R for any $F > 0$. But the figure 
prominently shows that for $F > 1$ and for small $T$, the mobility decreases instead
of increasing with temperature. This in itself is a novel feature. The above mentioned 
features result from a subtle combined effect of the periodic space dependent 
friction and the periodic potential in the presence of a constant applied force.
 The effect can be observed only when there is a phase difference between the
potential and the frictional profile; the phase difference makes the mobility asymmetric with respect to the reversal of the field $F$.
 The mobility shows many other interesting features as a function of other parameters, $F, \lambda$ and $\phi$. \cite{dan}

\section{Conclusion \hfill}

 We observe the occurrence of  stochastic resonance in the the mobility of an overdamped
Brownian particle in a sinusoidal potential tilted by a constant
force and subjected to a Gaussian white noise but, of course, in an inhomogeneous
 system with space-dependent friction coefficient.
 Thus the time dependent external oscillating field, which is generally
 considered as an essential ingredient for the observability of SR can be
 replaced by a constant force field concomitant with a space-dependent(periodic) friction coefficient of a spatially
 extended periodic system. We would like to point out here that the correct 
high friction Langevin equation in the space dependent frictional medium 
involves a multiplicative noise along with a temperature drift term \cite{Jay,Mah}. 

\section{ Acknowledgement \hfill}

M. C. M thanks the Institute of Physics, Bhubaneswar, for financial assistance and
hospitality. M. C. M and A. M. J acknowledge partial financial support from the \newline B. R. N. S project, D. A. E, India.
\newpage
\hspace{1.8in} \large REFERENCES \hfill

\newpage
  \hspace{1.5in} \large  FIGURE CAPTIONS.

\large
\vspace{0.5in}
\begin{flushleft}
Fig. 1.\hspace{0.3in} Mobility $\eta_{0}\mu$ as a function of temperature $T$ for $\phi = 0.9\pi$ and
$\lambda = 0.9$ for \\ \hspace{.8in}various values of F. The inset is inserted to highlight 
the peaks.
\vspace{0.1in}

Fig. 2. \hspace{0.3in}Mobility $\eta_{0}\mu$ versus $T$ for $\phi = 1.44\pi$, $\lambda = 0.9$ 
for various values of F. The inset \\ \hspace{.8in}highlights the peaks.
\vspace{0.1in}

Fig. 3. \hspace{0.3in}Mobility $\eta_{0}\mu$ versus $T$ for $\phi = 1.6\pi$, $\lambda = 0.9$ 
for various values of F. The inset \\ \hspace{.8in}highlights the minima.
\vspace{0.1in}
\end{flushleft}
\begin{figure}
  \centerline{\epsfbox{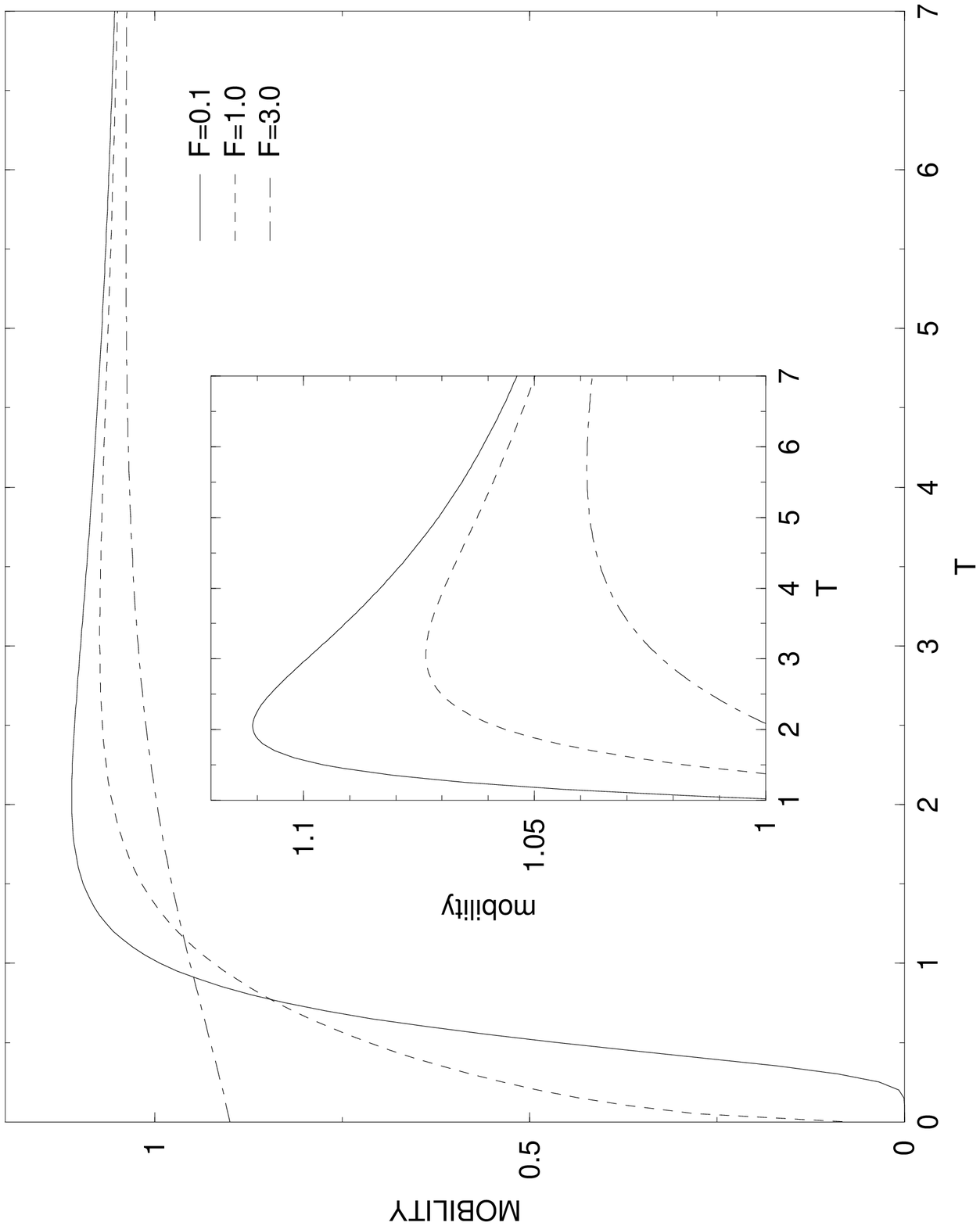}}
  \caption{}
\label{phase=0.45.eps}
\end{figure}

\begin{figure}
  \centerline{\epsfbox{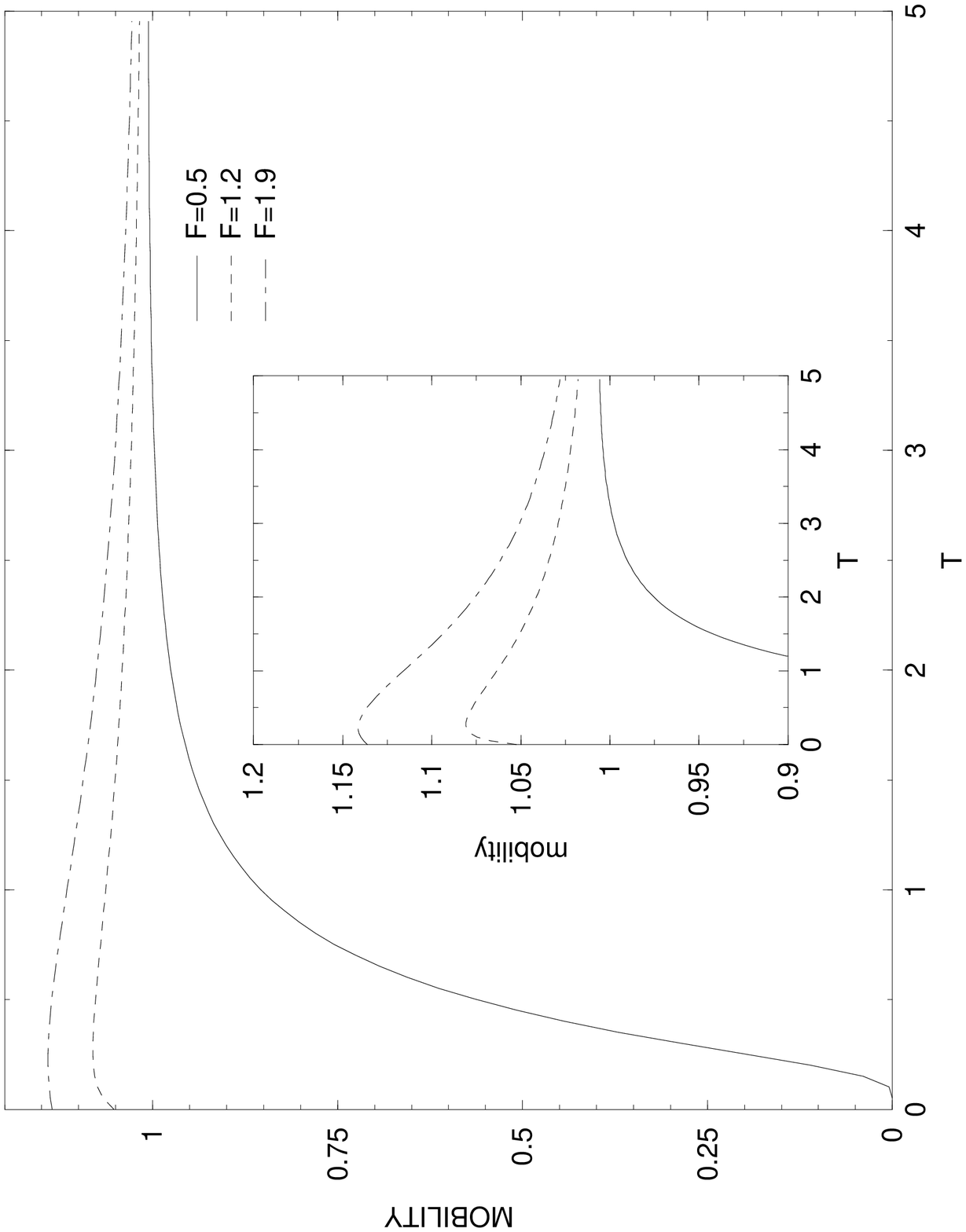}}
  \caption{}
\label{phase=.72.eps}
\end{figure}

\begin{figure}
  \centerline{\epsfbox{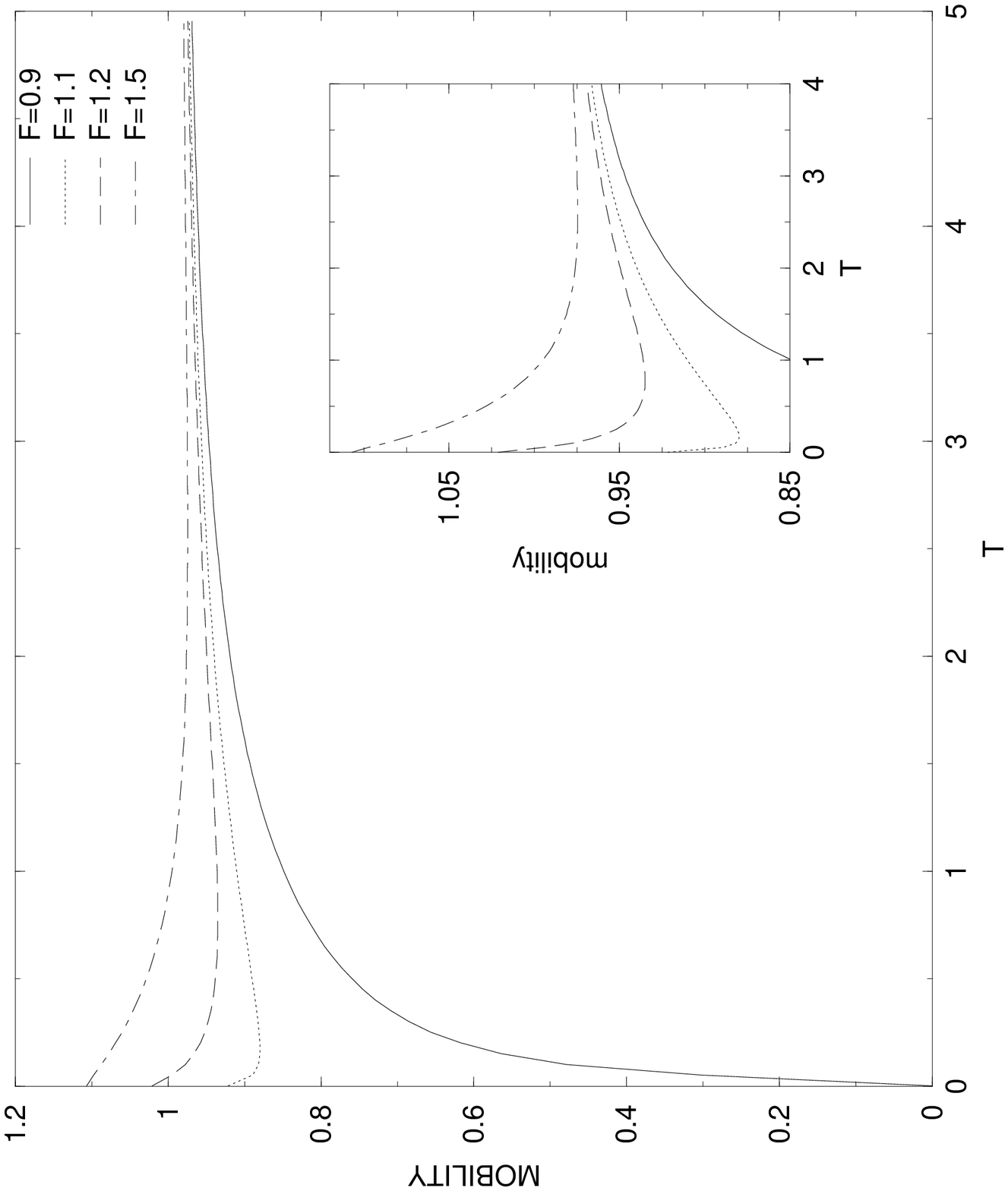}}
  \caption{}
\label{phase=0.8.eps}
\end{figure}
\end{document}